# Trajectory Privacy Protection Mechanism based on Social Attributes

Hua Wang

College of Information Science and Engineering, Hunan University,410082,ChangSha Chnia

Nest2018@hnu.edu.cn

**Abstract**.

The current trajectory privacy protection technology only considers the temporal and spatial attributes of trajectory data, but ignores the social attributes. However, there is an intrinsic relationship between social attributes and human activity trajectories, which brings new challenges to trajectory privacy protection, making existing trajectory privacy protection technologies unable to resist trajectory privacy attacks based on social attributes. To this end, this paper first studies the social privacy attack in the trajectory data, builds a social privacy attack model based on the fusion of "space-time" features, and reveals the internal impact of the spatial and temporal features in the trajectory data on social privacy leaks. -Anonymous algorithm and trajectory release privacy protection provide theoretical support. On this basis, integrate social attributes into trajectory privacy protection technology, design trajectory k-anonymity algorithm based on "space-time-social" three-dimensional mobile model, and construct trajectory data based on "space-time-social-semantic" multi-dimensional correlation Publish privacy-preserving models.

## 1 Introduction

The development of location positioning and wireless communication technology has promoted the popularization of mobile computing and location-based services (LBS, location-based services) [1]. In recent years, the service of LBS has gradually expanded from a single positioning/navigation to multiple vertical fields, including tourism, local search and information services, social networking and entertainment, leisure and fitness, family and personnel location services, mobile advertising, etc. And showing a trend of vigorous development [2]. The application scenarios of LBS cover people's daily life, and its industrial ecosystem has gradually matured.

However, while location-based services bring convenience and economic benefits to human life, they also cause many problems of data security and user privacy leakage [3]. Over the past few years, we've seen a lot of news related to privacy breaches about how companies are monetizing their users' data, and exposure of many practices that have sparked public outcry. For example, on July 11, 2014, News Channel exposed the issue of Apple's iPhone collecting users' location privacy, and reported that Apple records users' movement trajectories and location information in detail [4]. WeChat's "People Nearby" feature makes it easy for other users to locate a user's location, leading to an increasing number of crimes [5]. In 2017, a Quartz investigation revealed that Android phones have been collecting the addresses of nearby cell towers and sending that data to Google, even when users have disabled location services [6]. The consequence is that Google has access to data about an individual's location and their activities that goes well beyond consumers' reasonable privacy expectations.

In recent years, many such incidents have not only aroused consumers' attention to privacy, but also attracted great attention from governments of various countries. For example, the EU's new data privacy protection regulation GDPR [7], a series of laws and regulations such as "National Medium and Long-Term Science and Technology Development Plan" [8], "Cyber Security Law" [9] and "Several Issues Concerning the Application of Laws in the Handling of Criminal Cases of Infringement of Citizens' Personal Information" [10] all show that the state attaches great importance to the protection of personal privacy information.

In the actual use of location services, users usually upload their own location information at different locations and at different times in order to obtain services. Even if the location information is anonymized, it can still be restored to trajectory data by the attacker through association [11-13]. In addition to reflecting the sensitive information contained in the user's movement process (such as the sensitive location visited), the trajectory data usually contains more abundant spatio-temporal, semantic and even social information than a single location information. By analyzing the user's trajectory, the attacker can not only infer the user's home/work address, hobbies, health status, economic status, religious beliefs and other personal privacy information, but also obtain the user's behavior patterns, habits and social relationships and other higher-level privacy information, which in turn leads to the disclosure of personal privacy [14].

With the cross-integration of social networks and location services, mobile social networks have the largest number of active users in the LBS application market. In this context, the problem of social relationship privacy leakage in trajectory data is particularly significant. Social relations represent highly sensitive private information, which is closely related to users' social identities. In fact, mobile social network users are already aware of this threat and deliberately hide their social connections. For example, among New York Facebook users, the proportion of users who choose to hide their friends list increased from 17.2% in 2010 to 56.2% in 2011 [15]. However, there are still many users who have not realized that their trajectory data can reveal their social connections [16]. If the attacker has the trajectory data of the attack object but has no social relationship, the attacker usually uses the trajectory data to infer the potential social relationship between mobile users to mine more relevant sensitive information. For example, the US National Security Agency (NSA) is known to collect data on the location and travel habits of citizens in order to find unknown contacts of the target.

At present, there have been many works on social privacy attack methods based on trajectory data, but they only consider the characteristics of the spatial dimension, ignoring the impact of the characteristics of the time dimension on social privacy leakage. Moreover, the existing trajectory privacy protection algorithms that resist location inference attacks all ignore the influence of social relationships on human movement patterns, thus failing to effectively resist location inference attacks.

Existing social privacy inference attacks based on trajectory data only consider the characteristics of the spatial dimension of trajectory data, while ignoring the impact of the characteristics of the temporal dimension on social association inference. Recent studies have shown that ignoring the time background information of co-occurrence events will cause the deviation between the estimated results and the real results [17], which will increase the false positive rate of the estimated results. Therefore, it is necessary to study the impact of temporal features in trajectory data on social relationship inference, and the joint attack model combining time and space.

The trajectory synthesis technology in the existing location service lacks the evaluation and countermeasures for the dimension of social relationship when considering the two dimensions of time and space [18]. Recent studies have shown that attackers can use social relationships as auxiliary information to infringe on the user's track privacy, and the latest research has confirmed that a class of technologies based on social network deanonymization can break through the existing four A trajectory synthesis method based on spatio-temporal correlation. Such privacy risks bring new challenges to social privacy attacks and enhanced trajectory synthesis[19-20].

Therefore, this paper intends to break through the existing theoretical research framework of trajectory privacy attack and protection, and to study the problem of social privacy leakage in trajectory data and the corresponding trajectory privacy protection technology, aiming to reveal the impact of spatial, temporal and semantic features in trajectory data on social privacy. Intrinsic impact of leakage, and integrate the social attributes of trajectory data into the existing trajectory privacy protection technology, based on the deep learning model GAN, design a trajectory generation model that can capture multi-dimensional features, in order to achieve more practical and wider applicability. Trajectory generation technology provides an important basis. The research results of this paper have important theoretical significance and application value for further reducing the risk of user privacy leakage and promoting the sustainable, healthy and stable development and wider application of the location-based service industry.

## 2 Social Privacy Attack Model

The trajectory data sets used in the existing research are all location-based social network users' check-in data, and the user's social relationship network graph is used as the reference information for comparing the results of the attack model. However, the check-in data of social networks only records the time stamp of the check-in, and does not have the time information of the user's departure. Therefore, existing works only design measures of social relationship strength from the spatial dimension of co-visited locations. However, people's trajectory data naturally has time and space attributes, and ignoring the time background of user location records will cause deviations in attack results. It has been observed that many factors in the time dimension, such as the length of time users stay in the same location, the time when meeting events occur, and the time interval between consecutive meeting events, will affect the success rate of social privacy attacks. Therefore, how to extract features, which features to extract, and how to fuse heterogeneous features are prerequisites for correctly understanding the nature of social privacy leakage and designing effective trajectory privacy preservation methods.

### 2.1 Feature extraction

According to the co-location record set and co-occurrence record set, the social attribute features of user pairs in the spatial dimension and time dimension are respectively extracted and quantified into a mathematical model (measure) that can reflect the strength of social association. According to the experimental results of the previous data observation, it is proposed to define six metrics: frequency, popularity, diversity, time interval, co-occurrence duration, and holiday ratio.

$$J(l_{u_i}^j) = \sum_{i'=1}^{n_f} \sum_{j'=1}^{n_{i'}} D_{\alpha_d}(l_{u_i}^j, l_{u_{i'}}^{j'}) T_{\alpha_t}(t_{u_i}^j, t_{u_{i'}}^{j'})$$

### 2.2 Heterogeneous feature fusion model

It is proposed to fuse heterogeneous metrics based on the neural network model to improve the accuracy of social relationship inference. Taking the feature measures of space and time dimensions as input, through a three-layer feed-forward neural network trained according to the error backpropagation algorithm, the model is trained by minimizing the cross entropy between the real value and the estimated value to obtain the final Fusion results.

$$SI_{t_{n_s,s}}(u_i, u_{u'}) = \pi_1 e^{-\pi_2 \frac{\left\| l_{u_i,s}^{n_s}, c_{u_{i'}}^1 \right\|}{\left\| c_{u_{i'}}^1, c_{t_{n_s,s}} \right\|}}$$

## 3 Trajectory k-anonymity Algorithm

From the perspective of privacy attacks, the social attributes of trajectory data will be used by attackers as background knowledge for attacking other personal privacy of users. From the perspective of privacy protection, the social attributes of trajectory data can also be used to enhance trajectory privacy protection technology based on fake trajectories. Most of the existing works are based on heuristic privacy protection algorithms, but do not pay attention to the user's movement rules. Although a few research works focus on the spatiotemporal correlation of user movement, they ignore that the user's mobile habits are also affected by social relations. , so it is vulnerable to location speculation attacks based on social attributes. How to model the constraints of social attributes on users' mobile habits and embed them into the trajectory privacy protection framework based on spatio-temporal correlation constraints is the key to designing privacy protection technologies against location speculation attacks, which is expected to break through the current trajectory privacy Bottlenecks in conservation research.

$$I_{t_{n_s,s}}(u_i, u_{i'}) = \omega_s SI_{t_{n_s,s}}(u_i, u_{i'}) + \omega_t TI_{t_{n_s,s}}(u_i, u_{i'})$$

**3.1 Three-dimensional mobility model construction**

In addition to being restricted by time and space, user movement is also affected by social relationships. Therefore, it is proposed to model the user's movement trajectory in three-dimensional space (time dimension, space dimension and social dimension). The user's mobile behavior exhibits strong periodicity and random jumps. The locations that move randomly are called "social locations" and the locations that move periodically are called "non-social locations". The user's mobile behavior is mapped to the three dimensions of time behavior, spatial behavior and social behavior to establish a three-dimensional mobile model of the user at any time. Then, based on this 3D mobile model, the location points with the same characteristics as the target user's real trajectory are sampled in each dimension, and finally the trajectory is synthesized by using the sampled location points to hide the user's real trajectory, so as to resist location speculation attacks.

$$Pr\left[l_{u_i,\bar{s}}^{n_{\bar{s}}}\right] = \sum_{j=1}^{m} Pr\left[l_{u_i,\bar{s}}^{n_{\bar{s}}} | S(t_{n_{\bar{s}},\bar{s}}) = C_{u_i}^j\right] \times Pr\left[S(t_{n_{\bar{s}},\bar{s}}) = C_{u_i}^j\right]$$

$$= \sum_{j=1}^{m} N\left(U_{C_{u_i}^j}, \Sigma_{C_{u_i}^j}\right) Pr\left[S(t_{n_{\bar{s}},\bar{s}}) = C_{u_i}^j\right]$$

**3.2 Design of trajectory k-anonymity algorithm**

In order to protect the location privacy of the target user, the dummy trajectory used to hide the user's trajectory should exhibit similar social, temporal, and spatial behavior patterns to the target user's real trajectory. Therefore, it is proposed to conduct three-dimensional vertical segmentation of the user trajectory data, that is, to constrain the user's movement rules in three different dimensions of "time-space-social", and then synthesize anonymous data similar to the target user's trajectory in each dimension. track. At the same time, in order to ensure the usefulness of trajectory data, the trajectory participating in the anonymization also needs to have similar statistical characteristics with the trajectory of the target user. Based on the idea of differential privacy, (F,k,l)-differential sampling is proposed (F is a specific trajectory data analysis task, k is the privacy parameter of the target user, 0<l<1), so that participants in the anonymous synthetic trajectory Has similar statistical characteristics to user trajectories.

**4 Privacy-preserving trajectory publishing model**

The existing heuristic feature extraction and modeling methods can no longer meet the requirements of high efficiency and high applicability for trajectory release in current and future application scenarios. At present, deep learning has achieved remarkable success in the fields of image processing and natural language processing. Among them, the generative confrontation model GAN has excellent performance in feature extraction of various types of data. How to use GAN to efficiently learn the multi-dimensional features contained in the trajectory data and the correlation between the features to evaluate the similarity between the synthetic trajectory and the real trajectory, achieve the purpose of privacy protection while taking into account the usefulness of the data, and then satisfy the privacy protection of trajectory publishing. The requirements of efficiency and high applicability are the key issues to be solved by this paper.

The trajectory data can be converted into a sequence set representing the stay behavior, and the elements are position (x, y), start time t and duration d. Let K denote the maximum number of repeated stops per location that can be modeled in trajectory embeddings. Let M be the embedding matrix of trajectory data, then M(x,y,k)=(t,d) means the kth stay at time t, position (x,y), and stay duration d.

$$Pr(v(s)|\lambda) = \frac{\pi_\lambda N(v(s)|\mu_\lambda, \Sigma_\lambda)}{\sum_{\lambda' \in \Lambda} \pi_{\lambda'} N(v(s)|\mu_{\lambda'}, \Sigma_{\lambda'})}$$

$$\arg\max_{\theta \in \square^{|U \cup L| \times d_2}} \prod_{i \in U \cup L} \prod_{i' \in N(i)} \log \frac{1}{1+e^{-\theta(i) \cdot \theta(i')}} + \prod_{i \in U \cup L} \prod_{i' \in N(i)'} \log \frac{1}{1+e^{\theta(i) \cdot \theta(i')}}$$

## 5 Experiment

### 5.1 Experimental Setting

In the model training phase, the trajectory data is discretized and converted into a position map tensor. The discriminator takes as input the embedding matrix of the real trajectory and the embedding matrix of the synthetic trajectory, and is simultaneously trained by the generator with random noise. In the sampling phase, the position tensor is processed and converted into a position track. The synthetic trajectories are compared with the real trajectories using measures of space-time, semantic and social dimensions to ensure the similarity between the synthetic trajectories and the real trajectories. Adversarially trained nonparametric generative models are able to approximate and sample from joint distributions based on input features of complex data. Nonparametric generative models use a model, such as a deep learning network, to learn an objective function.

| ID | Start time | Start lat | Start lon | Stop time | Stop lat | Stop lon |
|---|---|---|---|---|---|---|
| 399387 | 16/09/2019 15:44:57 | 28.027098 | 112.973641 | 16/09/2019 15:50:11 | 28.032458 | 112.988596 |
| 399387 | 16/09/2019 16:13:49 | 28.032455 | 112.988666 | 16/09/2019 16:49:31 | 28.135861 | 113.049108 |

### 5.2 Experimental results and analysis

For the ability of the NPE evaluation model to maintain the decision structure, we first assess the overall performance of the method used to distinguish between positive and negative user decisions. The relationship between NPE and decision analysis is that NPE maintains the decision structure by identifying key factors and maximizing scale projection. The effectiveness of the NPE [12] in maintaining the decision-making structure depends directly on the quality of the key factors identified.

| Feature | Precision | Recall | F1-score |
|---|---|---|---|
| $f_{fre}$ | 0.701 | 0.628 | 0.801 |
| $f_{stay}$ | 0.804 | 0.804 | 0.804 |
| $f_{fre} + f_{pop}$ | 0.769 | 0.782 | 0.741 |
| Spatial features | 0.848 | 0.827 | 0.840 |
| Temporal features | 0.856 | 0.849 | 0.802 |
| $f_{fre} + f_{pop} + f_{stay}$ | 0.926 | 0.926 | 0.926 |
| All features | 0.957 | 0.957 | 0.957 |

Demonstrating the reliability of the semantic-aware representation module, Figure 1 shows the intermediate results of this module, and the impact of this module on attack performance. Figure 2 shows the mean vectors of Gaussian distributions for four visit purposes. Among them, purpose 1 represents entertainment venues, including attractions, amusement parks, etc. purpose 2 means shopping mall. purpose 3 indicates that it is used in residential areas. purpose 4 is the life service category, such as hospitals, communication business halls, etc. Figure 2 shows the performance comparison of the complete SEME scheme and the scheme "SEME-" (SEME without step1) without the semantic-aware representation module. The figure shows that the AUC value of SEME is higher than that of "SEME-", that is, by considering the semantic information (access purpose) of trajectory data, the attack performance is improved.

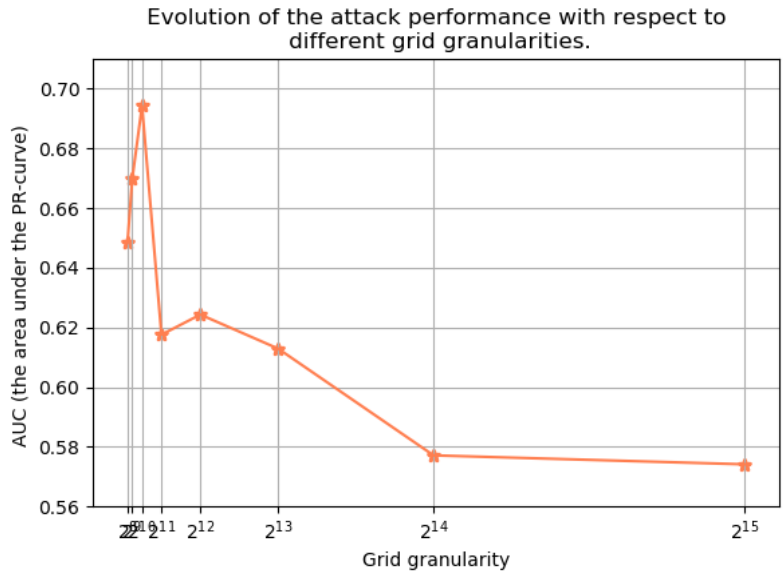

Figure2 Performance Impact of Semantic Awareness Module

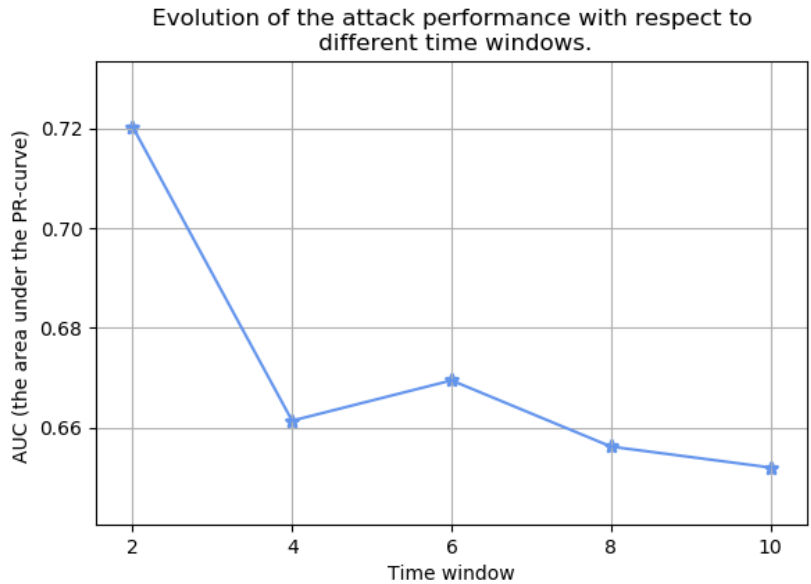

Figure3 Performance Impact of Modeling 3D Mobile Behavior

## 6 Conclusion

In this paper, the privacy protection framework when users initiate location-based service requests in the edge environment is carried out with the priority of user service quality, and its privacy protection approach and the way of evaluating the results returned by the edge server and cloud server are investigated. A privacy protection scheme based on pseudonymity and k-anonymity is designed, and RSA encryption algorithm is used to encrypt the user's private data. Two evaluation models are also proposed: NPE and POE, which combine privacy protection and related methods of destination prediction and evaluation, and can order the services returned by the server to the user to be more in line with the user's needs and can satisfy the user while safeguarding the user's privacy. Based on this, subsequent research directions can be to optimize the decision and response speed of task reproduction while protecting user privacy, and more reasonable resource allocation schemes need to be considered.

## 7 References

[1] H Jiang, S Jin, C Wang. Parameter-based data aggregation for statistical information extraction in wireless sensor networks[J]. IEEE Transactions on Vehicular Technology, 2010, 59(8): 3992-4001.

[2] Yennun Huang, Yih-Farn Chen, Rittwik Jana, Hongbo Jiang, Michael Rabinovich, Amy Reibman, Bin Wei, Zhen Xiao. Capacity analysis of MediaGrid: a P2P IPTV platform for fiber to the node (FTTN) networks[J]. IEEE Journal on Selected Areas in Communications, 2007, 25(1): 131-139.

[3] W Liu, D Wang, H Jiang, W Liu, C Wang. Approximate convex decomposition based

localization in wireless sensor networks[C]//2012 Proceedings IEEE INFOCOM. IEEE, 2012: 1853-1861.

[4] C Tian, H Jiang, X Liu, X Wang, W Liu, Y Wang. Tri-message: A lightweight time synchronization protocol for high latency and resource-constrained networks[C]//2009 IEEE International Conference on Communications. IEEE, 2009: 1-5.

[5] Y Huang, Z Xiao, D Wang, H Jiang, D Wu. Exploring individual travel patterns across private car trajectory data[J]. IEEE Transactions on Intelligent Transportation Systems, 2019, 21(12): 5036-5050.

[6] K Chen, C Wang, Z Yin, H Jiang, G Tan. Slide: Towards fast and accurate mobile fingerprinting for Wi-Fi indoor positioning systems[J]. IEEE Sensors Journal, 2017, 18(3): 1213-1223.

[7] H Huang, H Yin, G Min, H Jiang, J Zhang, Y Wu. Data-driven information plane in software-defined networking[J]. IEEE Communications Magazine, 2017, 55(6): 218-224.

[8] T Liu, JCS Lui, X Ma, H Jiang. Enabling relay-assisted D2D communication for cellular networks: Algorithm and protocols[J]. IEEE Internet of Things Journal, 2018, 5(4): 3136-3150.

[9] Jie Li, Fanzi Zeng, Zhu Xiao, Hongbo Jiang, Zhirun Zheng, Wenping Liu, Ju Ren. Drive2friends: Inferring social relationships from individual vehicle mobility data[J]. IEEE Internet of Things Journal, 2020, 7(6): 5116-5127.

[10] Z Xiao, D Xiao, V Havyarimana, H Jiang, D Liu, D Wang, F Zeng. Toward accurate vehicle state estimation under non-Gaussian noises[J]. IEEE Internet of Things Journal, 2019, 6(6): 10652-10664.

[11] P Zhang, X Chen, X Ma, Y Wu, H Jiang, D Fang, Z Tang, Y Ma. SmartMTra: Robust indoor trajectory tracing using smartphones[J]. IEEE Sensors Journal, 2017, 17(12): 3613-3624.

[12] S Wang, A Vasilakos, H Jiang, X Ma, W Liu, K Peng, B Liu, Y Dong. Energy efficient broadcasting using network coding aware protocol in wireless ad hoc network[C]//2011 IEEE International Conference on Communications (ICC). IEEE, 2011: 1-5.

[13] H Jiang, Z Ge, S Jin, J Wang. Network prefix-level traffic profiling: Characterizing, modeling, and evaluation[J]. Computer Networks, 2010, 54(18): 3327-3340.

[14] H Jiang, A Iyengar, E Nahum, W Segmuller, A Tantawi, CP Wright. Load balancing for

SIP server clusters[C]//IEEE INFOCOM 2009. IEEE, 2009: 2286-2294.

[15] H Jiang, P Zhao, C Wang. RobLoP: Towards robust privacy preserving against location dependent attacks in continuous LBS queries[J]. IEEE/ACM Transactions on Networking, 2018, 26(2): 1018-1032.

[16] H Jiang, J Cheng, D Wang, C Wang, G Tan. Continuous multi-dimensional top-k query processing in sensor networks[C]//2011 Proceedings IEEE INFOCOM. IEEE, 2011: 793-801.

[17] X Ma, H Wang, H Li, J Liu, H Jiang. Exploring sharing patterns for video recommendation on YouTube-like social media[J]. Multimedia Systems, 2014, 20(6): 675-691.

[18] H Jiang, W Liu, D Wang, C Tian, X Bai, X Liu, Y Wu, W Liu. CASE: Connectivity-based skeleton extraction in wireless sensor networks[C]//IEEE INFOCOM 2009. IEEE, 2009: 2916-2920.

[19] D Wang, J Fan, Z Xiao, H Jiang, H Chen, F Zeng, K Li. Stop-and-wait: Discover aggregation effect based on private car trajectory data[J]. IEEE transactions on intelligent transportation systems, 2018, 20(10): 3623-3633.